# Dissipative Relativistic Bohmian Mechanics

Roumen Tsekov

Department of Physical Chemistry, University of Sofia, 1164 Sofia, Bulgaria

It is shown that quantum entanglement is the only force able to maintain the fourth state of matter, possessing fixed shape at an arbitrary volume. Accordingly, a new relativistic Schrödinger equation is derived and transformed further to the relativistic Bohmian mechanics via the Madelung transformation. Three dissipative models are proposed as extensions of the quantum relativistic Hamilton-Jacobi equation. The corresponding dispersion relations are obtained.

Traditionally, the state of matter is recognized from its volume and shape properties. The solid state possesses both fixed volume and shape, while the liquid state maintains a fixed volume at variable shape. The gaseous state has both variable volume and shape, adapting them to fit the container. Usually the plasma is considered as the fourth state of matter. From its definition as a neutral mixture of charged particles, however, it follows that the traditional plasma is a gas with both variable volume and shape. Furthermore, ionic liquids (e.g. RTIL) and crystals (e.g. NaCl) could be considered as liquid and solid plasmas, respectively. The liquid and solid metals are definitely plasmas as well. The forgoing logic shows that the only possibility of the fourth state of matter is to possess variable volume at fixed shape. The reason for the different states of matter is the forces acting between the particles of matter. The thermal energy in solids is so small as compared to the potential interactions that the positions of the particles are firmly fixed and only small vibrations around them are present. That is why the solids maintain both fixed volume and shape. At higher temperature, the particles in liquids still cannot separate each other but they can move, allowing the liquids to flow. In gasses, the thermal energy is so high that the potential interactions between particles are negligible. In any case, the classical potentials decrease with increase of the distance between the particles and a very dilute matter is always an ideal gas (the Boyle's law). From this perspective the fourth state of matter, which does not possess a fixed volume, cannot be explained by the classical interaction potentials.

The ability to maintain an own shape even in a very dilute state requires interactions, which are not depressed by the distance. At present, the only known interaction, being independent from the distance between particles, is due to quantum mechanics. It is possible to generate quantum particles in such a way that the quantum state is defined only for the whole system. Thus, the quantum state of each particle depends on the states of the others, without presence of any classical potential interactions. This so-called quantum entanglement exists even if the particles are separated by a large distance. Therefore, the fourth state of matter could be an entangled one. The quantum entanglement is quantitatively described via the Bohm quantum potential.[1] It is latter recognized that the Bohm potential is an information potential[2] and represents the Fisher information force.[3] Due to the very close relationship between information and entropy,[4] the information forces are entropic and, hence, they differ from the usual potential interactions. The lack of volume restrictions suggests that the fourth state could probably be present at cosmological scale, which requires a relativistic treatment of the quantum problem.

In the beginning of the previous century Einstein's relativity and quantum mechanics have reformulated physics. The square root in the special relativity expression $E = \sqrt{m^2c^4 + p^2c^2}$ for the energy of a particle generates, however, some quantization problems and a usual way to get through is to consider its quadrate. Introducing the energy $\hat{E} \equiv i\hbar\partial_t$ and momentum $\hat{p} \equiv -i\hbar\nabla$ operators from quantum mechanics into $\hat{E}^2 = m^2c^4 + \hat{p}^2c^2$ yields the Klein-Gordon equation[5]

$$\Box\psi + (mc/\hbar)^2\psi = 0 \qquad (1)$$

where $\Box \equiv \partial_t^2/c^2 - \nabla^2$ is the d'Alembert operator. This partial differential equation describes scalar bosonic fields and reduces particularly to the wave equation $\Box\psi = 0$ for photons, since their rest mass $m \equiv 0$ is zero. The Klein-Gordon equation (1) suffers, however, serious problems with the probabilistic interpretation of the wave function $\psi$.[5] The reason for this could be an improper quantization of the rest mass energy, which must be persistent in space and time.

To resolve the problem one can introduce an energy operator, where the rest mass energy remains constant. Substituting $\hat{E} \equiv mc^2 + i\hbar\partial_t$ into the quadrate $\hat{E}^2 = m^2c^4 + \hat{p}^2c^2$ of the relativistic energy yields another fundamental equation for the relativistic quantum mechanics

$$\Box\psi + (2m/i\hbar)\partial_t\psi = 0 \qquad (2)$$

which reduces also to the wave equation for photons. As is seen, the wave function now is not a Lorentz invariant but $\psi$ is not an observable quantity. The physically relevant quantity is the probability density $\rho \equiv \psi\bar{\psi}$, which is invariant. One can easily recognize in Eq. (2) the relativistic Schrödinger equation, which can be rewritten in the alternative form

$$i\hbar\partial_t\psi = \hbar^2\Box\psi/2m + U\psi = \hat{H}\psi + \hbar^2\partial_t^2\psi/2mc^2 \qquad (3)$$

thus accounting for an external potential energy $U$ as well. The standard Hamiltonian operator reads $\hat{H} \equiv -\hbar^2 \nabla^2 / 2m + U$. A simple relationship $E_n = \sqrt{m^2 c^4 + 2m\varepsilon_n c^2}$ between the full relativistic $E_n$ and nonrelativistic $\varepsilon_n$ energy eigenvalues follows from Eq. (3), which resembles the Einstein expression. In the limit $c \to \infty$ Eq. (3) reduces naturally to the nonrelativistic Schrödinger equation, while the relativistic energy expands in series as $E_n = mc^2 + \varepsilon_n - \varepsilon_n^2 / 2mc^2 + \cdots$. To demonstrate that the relativistic Schrödinger equation (3) overcomes the probability problems of the Klein-Gordon equation (1) let us introduce the Madelung transformation of the complex wave function $\psi = \sqrt{\rho} \exp(iS/\hbar)$, where $S$ is the real quantum phase and $\rho$ is the local probability density. Thus, Eq. (3) reduces straightforward to the following two real equations

$$\partial_t \rho = \partial^\mu (\rho \partial_\mu S / m) \qquad \partial_t S - (\partial^\mu S)(\partial_\mu S)/2m + U + Q = 0 \qquad (4)$$

where $\partial_\mu$ is the standard 4-gradient operator with $\partial^\mu \partial_\mu = \Box$. The first equation is the continuity equation, while the second one is the relativistic quantum Hamilton-Jacobi equation. The latter differs from the classical analog[6] via the additional term $Q \equiv \hbar^2 \Box \sqrt{\rho} / 2m\sqrt{\rho}$, being the relativistic Bohm quantum potential.[7] As is seen, the latter can be strong even at vanishing matter density as required for the fourth state. Since the probability is conserved, $\int \rho d^3 x = 1$, the direct integration of the continuity equation leads to the following expression $\partial_t (\int \rho \partial_t S d^3 x) = 0$. It reflects the conservation of energy $E = -\int \rho \partial_t S d^3 x$, which is an integral of motion, independent of time.

The system of Eq. (4) defines the relativistic Bohmian mechanics. If one is interested in open quantum systems,[8] the relativistic Hamilton-Jacobi equation can be further extended to[9]

$$\partial_t S - (\partial^\mu S)(\partial_\mu S)/2m + U + Q = -\gamma S \qquad (5)$$

where the new term on the right-hand side describes the quantum phase decay with a collision frequency $\gamma$. If the latter is high enough, one can neglect the second nonlinear term and substituting $-\gamma S = \partial_t S + U + Q$ into the continuity Eq. (4) yields a relativistic quantum telegraph-like equation

$$\gamma \partial_t \rho = -\partial^\mu [\rho \partial_\mu (U + Q + \partial_t S)/m] = -\partial^\mu [\rho \partial_\mu (U + Q)/m] - \partial_t^2 \rho \qquad (6)$$

The last expression follows from the time derivative of the continuity equation, linearized on $S$ again. Since the Bohm quantum potential is a nonlinear function of the probability density, one can linearize further Eq. (6) on $\rho$ to obtain the following linear equation

$$\partial_t^2 \rho + \gamma \partial_t \rho + (\hbar/2m)^2 \square^2 \rho + \partial^\mu (\rho \partial_\mu U / m) = 0 \qquad (7)$$

In the case of a free particle ($U \equiv 0$) the time-space Fourier transformation of Eq. (7) provides the dispersion relation

$$-\omega^2 + i\omega\gamma + (\hbar/2m)^2 (k^2 - \omega^2/c^2)^2 = 0 \qquad (8)$$

At low frequency Eq. (8) simplifies to the imaginary nonrelativistic solution $\omega = i(\hbar k^2 / 2m)^2 / \gamma$, while at high $\omega$ the spectrum is modulated by the Zitterbewegung frequency $2mc^2/\hbar$. In general, Eq. (8) is a complex relationship between three important characteristic frequencies reflecting the collisions, Zitterbewegung and super-relativistic propagation with frequency $ck$.

Another dissipative model implies radiative friction,[10] where the corresponding relativistic Hamilton-Jacobi equation reads

$$\partial_t S - (\partial^\mu S)(\partial_\mu S)/2m + U + Q = \tau \partial_t^2 S \qquad (9)$$

If the characteristic time equals to $\tau = e^2 / 6\pi\varepsilon_0 mc^3$, for instance, the term on the right-hand side describes emission of photons. Omitting again the nonlinear term in Eq. (9) provides the rate of change of the quantum phase, $\tau \partial_t^2 S - \partial_t S = U + Q$. Thus, applying a time derivative twice on the continuity equation (4), neglecting the nonlinear $S$-terms and using the last equation yields

$$\partial_t^2 \rho - \tau \partial_t^3 \rho = -\partial^\mu [\rho \partial_\mu (U+Q)/m] \qquad (10)$$

One can linearize further Eq. (10) for low probability density gradients to obtain

$$\partial_t^2 \rho - \tau \partial_t^3 \rho + (\hbar/2m)^2 \square^2 \rho + \partial^\mu (\rho \partial_\mu U / m) = 0 \qquad (11)$$

In the case of a free particle ($U \equiv 0$) the time-space Fourier transformation of Eq. (11) provides the following dispersion relation

$$-\omega^2 + i\omega^3\tau + (\hbar/2m)^2(k^2 - \omega^2/c^2)^2 = 0 \qquad (12)$$

which simplifies at low and high frequency to $\omega^3 = i(\hbar k^2/2m)^2/\tau$ and $i\omega = \tau(2mc^2/\hbar)^2$, respectively. Comparing Eq. (12) and Eq. (8) unveils an effective friction coefficient $\gamma_\omega \equiv \tau\omega^2$.

The last dissipative model considered in the present paper implies diffusion of the quantum phase $S$, where the corresponding relativistic Hamilton-Jacobi equation reads[11]

$$\partial_t S - (\partial^\mu S)(\partial_\mu S)/2m + U + Q = D\nabla^2 S \qquad (13)$$

The new term describes reactive diffusion of $S$ with a diffusion constant $D$. Omitting again the nonlinear term in Eq. (13) provides the rate of change $\partial_t S = -U - Q + D\nabla^2 S$ of the quantum phase. Introducing it into the time derivative of the continuity Eq. (4), being linearized on $S$, yields

$$\partial_t^2 \rho = -\partial^\mu[\rho\partial_\mu(U+Q)/m] + D\partial^\mu(\rho\partial_\mu\nabla^2 S/m) \qquad (14)$$

One can linearize further Eq. (14) for low probability density gradients to obtain

$$\partial_t^2\rho - D\nabla^2\partial_t\rho + (\hbar/2m)^2\Box^2\rho + \partial^\mu(\rho\partial_\mu U/m) = 0 \qquad (15)$$

In the case of a free particle ($U \equiv 0$) the time-space Fourier transformation of Eq. (15) provides the dispersion relation

$$-\omega^2 + i\omega D k^2 + (\hbar/2m)^2(k^2 - \omega^2/c^2)^2 = 0 \qquad (16)$$

At low and high frequencies it simplifies to $\omega = i(\hbar k/2m)^2/D$ and $i\omega^3 = D(2mc^2 k/\hbar)^2$, respectively. Comparing Eq. (16) with Eq. (8) unveils an effective friction coefficient $\gamma_k \equiv Dk^2$. In the super-relativistic case ($\omega \simeq ck$) Eq. (16) reduces to simple diffusion, which is also the case of Eq. (12) with a diffusion constant $D = \tau c^2 = e^2/6\pi\varepsilon_0 mc$. Therefore, in the case of a complete relativistic treatment of the phase diffusion, the relativistic quantum Hamilton-Jacobi equation reads

$$\partial_t S - (\partial^\mu S)(\partial_\mu S)/2m + U + Q = -D\Box S \qquad (17)$$